\documentclass[aps,prd,10pt,twocolumn,superscriptaddress,showpacs,footinbib]{revtex4-1}
\pdfoutput=1
\usepackage{hyperref}
\usepackage{slashed}
\usepackage{graphicx}
\usepackage{enumerate}
\usepackage{amsfonts,amsmath,amssymb,bm,bbm}
\usepackage{color}
\usepackage{epsfig, subfigure}
\usepackage{setspace}
\usepackage{booktabs, tabularx} 
\usepackage{multirow}

\hypersetup{
    pdfnewwindow=true,      
    colorlinks=true,       
    linkcolor=blue,          
    citecolor=blue,        
    filecolor=blue,      
    urlcolor=blue        
}

\newcommand{\be}{\begin{equation}}  
\newcommand{\ee}{\end{equation}}
\newcommand{\bea}{\begin{eqnarray}}  
\newcommand{\eea}{\end{eqnarray}}

\newcommand{\nue}{\nu_e}

\newcommand{\nueb}{\bar{\nu}_e}

\begin{document}

\title{New Power to Measure Supernova $\nu_e$ with Large Liquid Scintillator Detectors}

\author{Ranjan Laha}
\affiliation{Kavli Institute for Particle Astrophysics and Cosmology, Stanford University, SLAC National Accelerator Laboratory, Menlo Park, CA 94025, USA}

\author{John F. Beacom}
\affiliation{Center for Cosmology and AstroParticle Physics (CCAPP), Ohio State University, Columbus, OH 43210, USA}
\affiliation{Department of Physics, Ohio State University, Columbus, OH 43210, USA}
\affiliation{Department of Astronomy, Ohio State University, Columbus, OH 43210, USA}

\author{Sanjib Kumar Agarwalla}
\affiliation{Institute of Physics, Sachivalaya Marg, Sainik School Post, Bhubaneswar 751005, India\\
{\tt rlaha@stanford.edu, beacom.7@osu.edu, sanjib@iopb.res.in} \smallskip}

\date{December 29, 2014}

\begin{abstract}
We examine the prospects for detecting supernova $\nu_e$ in JUNO, RENO-50, LENA, or other approved or proposed large liquid scintillator detectors.  The main detection channels for supernova $\nu_e$ in a liquid scintillator are its elastic scattering with electrons and its charged-current interaction with the $^{12}$C nucleus.  In existing scintillator detectors, the numbers of events from these interactions are too small to be very useful.  However, at the 20-kton scale planned for the new detectors, these channels become powerful tools for probing the $\nu_e$ emission.  We find that the $\nu_e$ spectrum can be well measured, to better than $\sim 40\%$ precision for the total energy and better than $\sim 25\%$ precision for the average energy.  This is adequate to distinguish even close average energies, e.g., 11 MeV and 14 MeV, which will test the predictions of supernova models.  In addition, it will help set constraints on neutrino mixing effects in supernovae by testing non-thermal spectra.  Without such large liquid scintillator detectors (or Super-Kamiokande with added gadolinium, which has similar capabilities), supernova $\nu_e$ will be measured poorly, holding back progress on understanding supernovae, neutrinos, and possible new physics.
\end{abstract}

\maketitle

\section{Introduction}
\label{sec:introduction}

If a core-collapse supernova in the Milky Way appears soon, will the neutrino detectors be ready?  Yes, in the sense that the supernova will not be missed, as there are many independent detectors.  But would this data be complete enough to answer pressing questions?  The answer is no, because not all flavors of neutrinos and antineutrinos will be measured well.  We might have to wait a few decades more to get the answers.

Only with all six flavors --- expected to be emitted with comparable total energies, but with different spectra and time profiles --- can we measure the combined neutrino emission, which reveals the change in gravitational binding energy of the stellar core as well as the effects of any novel energy-loss processes~\cite{Mueller:2014oia,Wongwathanarat:2014yda}.  And only with all six flavors can we test how this energy is apportioned, which reveals the density and neutron-to-proton ratio of the collapsing core as well as the effects of neutrino mixing in extreme conditions~\cite{Dighe:1999bi,Duan:2010bg,Dasgupta:2010gr}.

The Super-Kamiokande detector will detect $\bar{\nu}_e$ with $\sim 10^4$ events, and will be able to reconstruct its spectrum precisely due to the tight connection between observable energy and neutrino energy in $\bar{\nu}_e \,+\, p \rightarrow e^+ \,+\, n$ interactions with free protons~\cite{Vogel:1999zy,Strumia:2003zx}.  The IceCube detector, though it cannot measure individual events or spectra, will measure the time profile of the $\bar{\nu}_e$ flux to high precision~\cite{Abbasi:2011ss}.  The utility of these precise measurements is limited by how well the other flavors can be measured.

The spectra and time profiles of $\nu_\mu$, $\nu_\tau$, $\bar{\nu}_\mu$, and $\bar{\nu}_\tau$ --- generically called $\nu_x$ --- are expected to be very similar to each other, which considerably simplifies the detection problem.  These flavors can be measured reasonably well, with $\sim 10^2$ events, in scintillator detectors (KamLAND, Borexino, and soon SNO+) through the $\nu + p \rightarrow \nu + p$ channel, which has good spectral fidelity for the high-energy part of the spectrum~\cite{Beacom:2002hs,Dasgupta:2011wg}.

The missing link is sufficient sensitivity to $\nu_e$.  Interactions with electrons are suppressed by the small electron mass, and interactions with neutrons are suppressed by nuclear binding effects.  Though Super-Kamiokande would have $\sim 10^2$ events caused by $\nu_e$, it is difficult to isolate these from other channels, and no other existing detectors would have appreciable numbers of events.

In a previous paper by two of us, we showed how this situation could be significantly improved if Super-Kamiokande adds dissolved gadolinium to enable neutron detection~\cite{Laha:2013hva}.  This would lead to clean identification of the dominant $\bar{\nu}_e + p \rightarrow e^+ + n$ events, making it easier to separate the $\nu_e$ events on electrons and nuclei.  We showed that the total energy emitted in $\nu_e$ and their average energy could each be measured to $\sim 20\%$ precision.

Here we examine how well $\nu_e$ could be measured in large liquid scintillator detectors --- those that are comparable to the size of Super-Kamiokande and more than an order of magnitude larger than existing scintillator detectors.  The JUNO (also known as Daya Bay II)~\cite{He:2014zwa,Li:2014qca} detector is already approved, and the RENO-50~\cite{Kim:2014rfa}, LENA~\cite{Wurm:2011zn} detectors are under consideration.  These detectors would have yields of $\bar{\nu}_e \,+\, p \rightarrow e^+ \,+\, n$ events comparable to that of Super-Kamiokande.  Importantly, they would have much larger yields of $\nu + p \rightarrow \nu + p$ events than existing detectors.  Most importantly, they would have newly powerful sensitivity to $\nu_e$ through interactions with electrons and nuclei.  Our presentation below closely follows that of Ref.~\cite{Laha:2013hva}, to make it easier to compare results, but fewer details are given here.

In Sec.~\ref{sec:Neutrino spectra and Gd_in_liquid_scintillator}, we describe how supernova $\nu_e$ can be detected in liquid scintillator detectors.  In Sec.~\ref{sec:SN nue detection_in_liquid_scintillator}, we estimate how well the parameters of the incident $\nu_e$ spectrum can be determined.  We conclude in Sec.~\ref{sec:conclusion_in_liquid_scintillator}.

\section{Supernova Neutrino Detection}
\label{sec:Neutrino spectra and Gd_in_liquid_scintillator}

We discuss the neutrino spectra expected from a supernova, the neutrino detection channels in liquid scintillator detectors, the likely experimental realities of large liquid scintillator detectors, and our proposed strategies to isolate $\nu_e$.

\subsection{Supernova Neutrino Spectra}

We assume that the total energy in neutrinos emitted by the supernova is $3 \times 10^{53}$ erg and that this is equally divided between all six active flavors of neutrinos and antineutrinos.  We take the distance to a typical Milky Way supernova to be 10 kpc~\cite{Adams:2013ana}.

For the spectra, we assume a (normalized) modified Maxwell-Boltzmann form~\cite{Keil:2002in,Tamborra:2012ac}
\begin{equation}
f(E_\nu)=\frac{128}{3}\frac{E_\nu ^3}{\langle E_\nu \rangle ^4} \,{\rm exp} \left({-\frac{4E_\nu}{\langle E_\nu \rangle}}\right)\, ,
\label{eq:normalised spectra}
\end{equation} 
where $E_\nu$ and $\langle E_\nu \rangle$ are the neutrino energy and average energy.  Compared to a regular Maxwell-Boltzmann form, this has somewhat fewer neutrinos at high energies, so our choice is conservative.  Typical average energies for the initial neutrino spectra from numerical supernova models are $\langle E_{\nu_e} \rangle \approx$ 11 -- 12 MeV, $\langle E_{\bar{\nu}_e} \rangle \approx$ 14 -- 15 MeV, and $\langle E_{\nu_x} \rangle \approx$ 15 -- 18 MeV.

Due to neutrino mixing in the supernova~\cite{Dighe:1999bi,Dasgupta:2007ws,Dasgupta:2008cd,Dighe:2008dq,Dasgupta:2009mg,Dasgupta:2010ae,Dasgupta:2010cd,Dasgupta:2010gr,Duan:2010bg,Duan:2010af,Friedland:2010sc,Pehlivan:2011hp,Sarikas:2011am,Cherry:2012zw,Borriello:2013tha,Chakraborty:2014nma} or in the Earth~\cite{Lunardini:2001pb,Dighe:2003jg,Dighe:2003vm,Borriello:2012zc}, the $\nu_e$ (or $\bar{\nu}_e$) spectrum could be modified (effectively made hotter) by mixing.  As in Ref.~\cite{Laha:2013hva}, we focus on scenarios where the expected temperature hierarchy occurs and where we seek to determine if the $\nu_e$ spectrum is affected by mixing or not; other scenarios can be tested separately.

\subsection{Detectable Neutrino Interactions}

The neutrino detection channels in a liquid scintillator detector are listed in~\cite{Cadonati:2000kq, Wurm:2011zn, Machado:2012ee, Scholberg:2012id, Lujan-Peschard:2014lta}.  Liquid scintillator detectors can detect electrons, positrons, photons and non-relativistic protons with near-perfect efficiency above a low energy threshold.  The detectable energy of a positron is its kinetic energy plus the energy deposited during annihilation, 2$m_e$, whereas the detachable energy for an electron is just its kinetic energy.  Due to the large number of photoelectrons produced per MeV, the energy and position resolution is excellent.  Neutrons can be detected with high efficiency via their radiative captures on protons and (rarely) carbon, as discussed below.

Electron antineutrinos $\nueb$ can be detected via the inverse beta interaction with free protons,  $\bar{\nu}_e \,+ \,p \rightarrow e^+ \,+ \,n$~\cite{Vogel:1999zy,Strumia:2003zx}.  The 2.2 MeV photon that results from neutron capture on protons is routinely detected in liquid scintillator detectors like KamLAND~\cite{Tolich:2011zz}, Double Chooz~\cite{Abe:2013sxa}, Daya-Bay~\cite{An:2014ehw}, and RENO~\cite{RENO:Neutrino2014}.  The double coincidence signal of $e^+$ and $n$ means that these events can be individually identified.   In water Cherenkov detector like Super-Kamiokande, Gadolinium loading would be required to unambiguously detect this interaction~\cite{Beacom:2003nk}.  

For the elastic scattering of neutrinos on electrons, we follow the discussion in~\cite{Laha:2013hva}, with the important difference that liquid scintillator detectors have no directionality.  The angular cut that can be employed in water Cherenkov detectors to suppress backgrounds is not available in liquid scintillator detectors.

All flavors of neutrinos elastically free protons via the neutral-current interaction.  The recoil energy of the scattered proton varies from zero to a maximum value that depends on the square of the incident neutrino energy~\cite{Beacom:2002hs,Dasgupta:2011wg}.  To account for quenching of the proton energy, we have taken the Birk's constant to be 0.01 cm/MeV, an indicative value that is similar to the measurements in~\cite{vonKrosigk:2013sa}.  The quenching factor in the detector will depend on the scintillator properties.  Due to the threshold of a liquid scintillator detector during supernova burst ($\sim$ 0.2 MeV), this interaction is only sensitive to the neutrino flavors with the highest average energies~\cite{Beacom:2002hs,Dasgupta:2011wg}, and the proton recoil spectrum can be used to reconstruct this neutrino spectrum~\cite{Dasgupta:2011wg}.

There are also important interactions of neutrinos with carbon nuclei.  Neutrinos interactions via neutral current can also excite the $J^\pi, T = 1^+, 1$ state in $^{12}$C, which then decays spontaneously to the ground state via the emission of a 15.11 MeV gamma-ray photon~\cite{Fukugita:1988hg, Kolbe:1995af, Engel:1996zt, Kolbe:1999au,Hayes:1999ew, Volpe:2000zn}.  Electron neutrinos $\nu_e$ interact with $^{12}$C to produce unstable $^{12}$N$_{\rm g.s.}$ in its ground state: 
\begin{eqnarray}
\nu_e \,+\, ^{12}{\rm C} \,\rightarrow \,^{12}{\rm N}_{\rm g.s.}  \,+ \, e^- \,.
\end{eqnarray}
The $^{12}$N$_{\rm g.s.}$ decays with a half-life of 11 msec: 
\begin{eqnarray}
^{12}{\rm N}_{\rm g.s.} \,\rightarrow \, ^{12}{\rm C} \,+\, e^+ \,+\, \nu_e \,.  
\end{eqnarray}
The maximum kinetic energy of the positron from the $^{12}{\rm N}_{\rm g.s.}$ decay is $\approx$ 16.8 MeV.  Similarly, electron antineutrinos $\bar{\nu}_e$ interact with $^{12}$C to produce unstable $^{12}$B$_{\rm g.s.}$: 
\begin{eqnarray}
\bar{\nu}_e \,+\, ^{12}{\rm C} \,\rightarrow\, ^{12}{\rm B}_{\rm g.s.} \,+\, e^+ \,.
\end{eqnarray}
The $^{12}$B$_{\rm g.s.}$ nuclei decays with a half-life of 20.2 msec:
\begin{eqnarray}
^{12}{\rm B}_{\rm g.s.} \,\rightarrow \, ^{12}{\rm C} \,+ \,e^- \,+\, \bar{\nu}_e \,.
\end{eqnarray}
The maximum kinetic energy of the electron from the $^{12}{\rm B}_{\rm g.s.}$ decay is $\approx$ 12.9 MeV.  Both types of charged-currrent events can be identified by the time and space coincidence of the scattering followed by a decay, though it is difficult to distinguish the two channels.  The cross section of the $\nu_e$ on $^{12}$C in the relevant energy range has been measured by the LSND collaboration~\cite{Auerbach:2001hz} and is in good agreement with theoretical calculations~\cite{Fukugita:1988hg, Kolbe:1995af, Engel:1996zt, Kolbe:1999au,Hayes:1999ew, Volpe:2000zn}, as expected because these cross sections depend primarily on the measured $^{12}$N$_{\rm g.s.}$ and $^{12}$B$_{\rm g.s.}$ lifetimes.  For these interactions, we use the cross sections tabulated in~\cite{Fukugita:1988hg}.  

In Borexino and Double Chooz, it has been shown that positrons can be separated from electrons via pulse shape discrimination, due to the deposited annihilation energy of the positron~\cite{Kino,2011PhRvC..83a5504F,Bellini:2013lnn,Abe:2014uba}.  The positron is detected via its annihilation energy after ortho-positronium formation (fraction $\sim$ 45\%), as this state has a long enough lifetime for detection with present technology.  Although this has not yet been demonstrated in KamLAND, it is expected that future large liquid scintillator detectors will have this capability.  Para-positronium has a lifetime which is three orders of magnitude shorter than ortho-positronium~\cite{Abe:2014uba}, so it is unrealistic to assume that there will be perfect separation of positrons from electrons.  However, even with perfect separation, the contours we calculate below would only improve by a factor of $\sim$ $\sqrt{2}$.

There are also neutrino interactions with $^{12}$C that are inelastic in the sense of emitted final-state nucleons; we neglect such channels.  The neutral-current interaction can cause $\nu \, + \, ^{12}$C $\rightarrow$ $\nu' \,+\, p \,+\, ^{11}$B.  The total number of interactions in a 20 kton detector varies from 3 to 31 for neutrino average energies of 12 MeV to 18 MeV.  Although the proton recoil spectrum for this interaction extends to a higher energy than due to neutrino-proton scattering~\cite{Lujan-Peschard:2014lta}, it will be difficult to detect them due to the low number of events.  Charged-current interactions can lead to particle-unbound excited states of $^{12}$N and $^{12}$B~\cite{Yoshida:2008zb}.  The largest yields are from $\nu_e$ + $^{12}$C $\rightarrow$ $e^-$ + $p$ + $^{11}$C and $\bar{\nu}_e$ + $^{12}$C $\rightarrow$ $e^+$ + $n$ + $^{11}$B.  Compared to interactions to the ground states, the number of interactions in a 20 kton liquid scintillator detector is between $\sim$ 10\% and $\sim$ 40\%, for incoming neutrino average energies of 12 MeV to 18 MeV respectively.   It may be possible to tag the $\nu_e$ events by the decays of $^{11}$C, as has been demonstrated in Borexino~\cite{Bellini:2013lnn}.  The $\bar{\nu}_e$ events will be hidden by the large yield of inverse beta interaction on free protons. 

Liquid scintillator detectors are also sensitive to neutrino interactions on rare $^{13}$C~\cite{Ianni:2005ki,Suzuki:2012aa}.  For $\nu_e$, there is a low threshold of 2.2 MeV, and transitions to the ground and 3.5 MeV excited state of $^{13}$C are important.  Using the cross sections calculated in~\cite{Suzuki:2012aa}, we find that the number of interactions expected in a 20 kton liquid scintillator detector is $\sim$ 10 -- 20, depending on the neutrino average energy.  Elaborate consideration of the background is needed to discover this signal~\cite{Ianni:2005ki}.  Due to the small number of events, we neglect it.

\subsection{Detector properties}

Large liquid scintillator detectors are being planned for a variety of physics reasons.  These include determination of the neutrino mass hierarchy, precision measurement of neutrino parameters, detection of supernova neutrinos, solar neutrinos, geoneutrinos, sterile neutrinos, atmospheric neutrinos, nucleon decay, and many other exotic searches~\cite{Wurm:2011zn,Machado:2012ee,Li:2014qca,RENO-50:2013,Bakhti:2013ora,Bakhti:2014pva,Park:2014sja,Mollenberg:2014pwa,Mollenberg:2014mfa,Kim:2014rfa,He:2014zwa}.

Due to its main design goal of detecting the neutrino mass hierarchy, a detector like JUNO~\cite{Li:2014qca,He:2014zwa} or RENO-50~\cite{Kim:2014rfa} has very specific features.  JUNO will have an inner volume of 20 kton and RENO-50 is being designed to have an inner volume of 18 kton.  We assume that the fiducial volume is 20 kton for supernova detection.  Although the fiducial volume of both the detectors will be somewhat smaller than the inner volume,  due to the short duration and lower backgrounds during a supernova burst, the fiducial volume during a supernova neutrino search can be almost as large as the inner volume.  LENA will have a much larger target mass of liquid scintillator, $\sim$ 50 kton~\cite{Wurm:2011zn}, and so the precision of our results would improve by a factor of $\sim$ $\sqrt{2.5}$ $\approx$ 1.6.

The JUNO liquid scintillator will be primarily linear alkyl benzene (C$_6$H$_5$C$_{12}$H$_{25}$)~\cite{Li:2014qca,He:2014zwa}.  The energy resolution of both JUNO and RENO-50 is expected to be $\sim$ 3\%/$\sqrt{E/{\rm MeV}}$~\cite{Li:2014qca}.  We neglect the impact of energy resolution in this work except in the case of the 15.11 MeV monochromatic photon.  Due to the intrinsic width of the detectable signals, the effect of the energy resolution can be neglected otherwise.

JUNO will not have added gadolinium, in order to achieve lower radioactivity level and higher transparency~\cite{He:2014zwa}.  In spite of this, JUNO is expected to have a near-perfect efficiency in detecting neutrons.  The mean lifetime of neutron capture on protons depend on the scintillator, e.g., it is $\sim$ 207 $\mu$sec in KamLAND~\cite{Collaboration:2011jza} and $\sim$ 260 $\mu$sec in Borexino~\cite{Bellini:2013pxa}.  By using a time cut of 0.5 $\mu$sec to 1000 $\mu$sec, nearly all neutrons are detected in KamLAND.  Because detector backgrounds can be neglected during the short time of a supernova burst, a long time cut can be used for neutron detection.

Neutron capture on protons yield a 2.2 MeV gamma-ray photon.  In addition to free protons, about 1\% of the neutrons will also be captured on carbon which yields a 4.9 MeV photon~\cite{Bellini:2013pxa}.  The width of these monochromatic photon energies is expected to be 0.044 MeV and 0.066 MeV respectively in JUNO and RENO-50.  KamLAND employs an energy cut of 1.8 MeV -- 2.6 MeV to detect all the neutron capture events on protons~\cite{Collaboration:2011jza}.  Due to the superior energy resolution of JUNO, an energy cut of 1.9 MeV -- 2.5 MeV will be sufficient.  Similarly an energy cut of 4.5 MeV -- 5.3 MeV will help in detecting all the neutron capture events on carbon.

In a liquid scintillator detector, the spatial cut for neutron capture on protons is mostly driven by the absorption length of the resulting 2.2 MeV photons and the vertex resolution~\cite{Collaboration:2011jza}.  A spatial cut of 1.6 m is used in KamLAND to achieve near perfect capture efficiency from this selection~\cite{Detwiler:2005hj,Dwyer:2007kra}.  The position resolution in JUNO or RENO-50 is expected to be much better and hence such a spatial will help capture all the neutron capture events on protons.  A different spatial cut is required to capture all the neutron capture events on carbon and we expect that these spatial cuts will ensure detection of all the neutron capture events.  The large volume of these detectors imply that the fraction of neutrons leaking out will be negligible.  

\begin{table}[b]
\caption{Expected numbers of events in a 20 kton liquid scintillator detector for a Galactic supernova for different values of the neutrino average energy.  The total energy is assumed to be $3 \times 10^{53}$ erg, divided equally among all flavors, at distance of 10 kpc.  The detection threshold during a burst is assumed to be $T_{\rm obs} = 0.2$ MeV.  For neutral current interactions, the numbers of events are for one flavor of $\nu$ or $\bar{\nu}$.}
\setlength{\extrarowheight}{4pt} 
\begin{ruledtabular}
\begin{spacing}{1.1}
\begin{tabular}{lccc}
Detection channel & 12 MeV & 15 MeV & 18 MeV \\ 
\hline
$\bar{\nu}_e \,+ \,p \rightarrow e^+ \,+ \,n$ & 3898 & 4857 & 5727 \\
\hline
$\nu$ \,+ \,$p$ $\rightarrow$ $\nu$ \,+ \,$p$ & 50 & 139 & 236\\
$\bar{\nu}$ \,+ \,$p$ $\rightarrow$ $\bar{\nu}$ \,+ \,$p$ & 50 & 130 & 236\\
\hline
$\nu_e \,+ \,e^- \rightarrow \nu_e \,+ \,e^-$ & 159 & 160 & 160 \\
$\bar{\nu}_e \,+ \,e^- \rightarrow \bar{\nu}_e \,+ \,e^-$ & 65 & 66 & 67 \\
$\nu_x \,+ \,e^- \rightarrow \nu_x \,+ \,e^-$ & 26 & 27 & 27 \\
$\bar{\nu}_x \,+ \,e^- \rightarrow \bar{\nu}_x \,+ \,e^-$ & 23 & 23 & 23 \\
\hline
$\nu_e \, + \, ^{12}$C $\rightarrow e^- \,+ \,^{12}$N$_{\rm g.s.}$ & 44 & 114 & 214 \\
$\bar{\nu}_e \, + \, ^{12}$C $\rightarrow e^+ \,+ \,^{12}$B$_{\rm g.s.}$ & 49 & 107 & 177 \\
$\nu \, + \,^{12}$C $\rightarrow \nu ' \,+ \,^{12}$C$^*$ (15.11) & 26 & 60 & 104 \\
$\bar{\nu} \,+ \,^{12}$C $\rightarrow \bar{\nu} ' \,+ \,^{12}$C$^*$ (15.11) & 24 & 56 & 95 \\
\end{tabular}
\end{spacing}
\end{ruledtabular}
\label{tab:yields}
\end{table}

We expect that the overall neutron capture efficiency in future large liquid scintillator detector to be 100\%.  KamLAND already has a $\sim$ 95\% efficiency of neutron capture on protons in their search for electron antineutrino from reactors~\cite{Abe:2008aa}.  A future liquid scintillator detector like JUNO or RENO-50 will be at a deeper site~\cite{Li:2014qca,RENO-50:2013} and hence will have lower background induced by muons~\cite{Kudryavtsev:2003aua, Arslan:2013bca}.  Since the lifetime of $^{12}$N$_{\rm g.s.}$ is very different from the neutron capture lifetime on protons, the impact of neutron detection inefficiency will be on the constraints from $\nu_e$ + $e^-$ events.  We will show the impact of 90\%, 95\%, 99\% and 100\% neutron detection efficiency on $\nu_e$ + e$^-$ events.  Subsequently we will assume 100\% neutron capture efficiency throughout the work. 

The detection threshold is determined by the radioactive background in the liquid scintillator detector and the surrounding rock.  The energy region below $\sim$ 0.2 MeV is dominated by $\beta$ decays of $^{14}$C nuclei.  Pulse shape discrimination can be used to reduce this background but it might still be high compared to neutrino interaction rates~\cite{Bellini:2009jr, Bellini:2014uqa}.  The background caused by the $\alpha$ decay of $^{210}$Po in the energy range 0.2 -- 0.5 MeV can be reduced to manageable levels~\cite{Villante:2011zh}.  Most of the $\nu \,+\, p$ elastic scattering events are in the energy range 0.2 -- 2 MeV.

The differential rate of the number of neutrino interactions in the different detectable channels are calculated following the discussion in Ref.~\cite{Laha:2013hva}.  Table~\ref{tab:yields} shows the expected number of events in a liquid scintillator detector with a fiducial volume of 20 kton when different average energies of the neutrino flavors are assumed.  From the table, it is clear that $\nu_e$ interactions on electrons are the largest in number among electron scattering interactions.  The rate of $\nu \,+\, e^-$ scattering events has little dependence on the average energy of the neutrino spectrum.  The steep energy dependence of the neutrino interactions on carbon is also evident from this table.  These points imply that we will obtain the best constraint on the $\nu_e$ average energy by using the double coincidence signal of $\nu_e$ interaction on $^{12}$C, whereas the electron scattering events will typically provide a better constraint on the total energy emitted in that particular flavor.

These signals can be divided into three broad categories depending on their temporal characteristics.  The elastic scattering of neutrinos with electrons are a single-signal event in which the final state electron is detected.  The neutral current scattering with the $^{12}$C nuclei which produces the 15.11 MeV monochromatic gamma-ray photon is also a single-signal event.  The inverse beta decay interaction is a double coincidence signal with a characteristic time of $200 \, \mu$s.  The charged current interaction of $\nu_e$ and $\bar{\nu}_e$ on $^{12}$C is also a double coincidence signal event in which the emitted charged leptons are separated by the time given by the half-life of the excited nucleus which is 11 or 20 msec.

\subsection{Detection Strategy}

\begin{figure*}
\centering
\includegraphics[angle=0.0,width=0.48\textwidth]{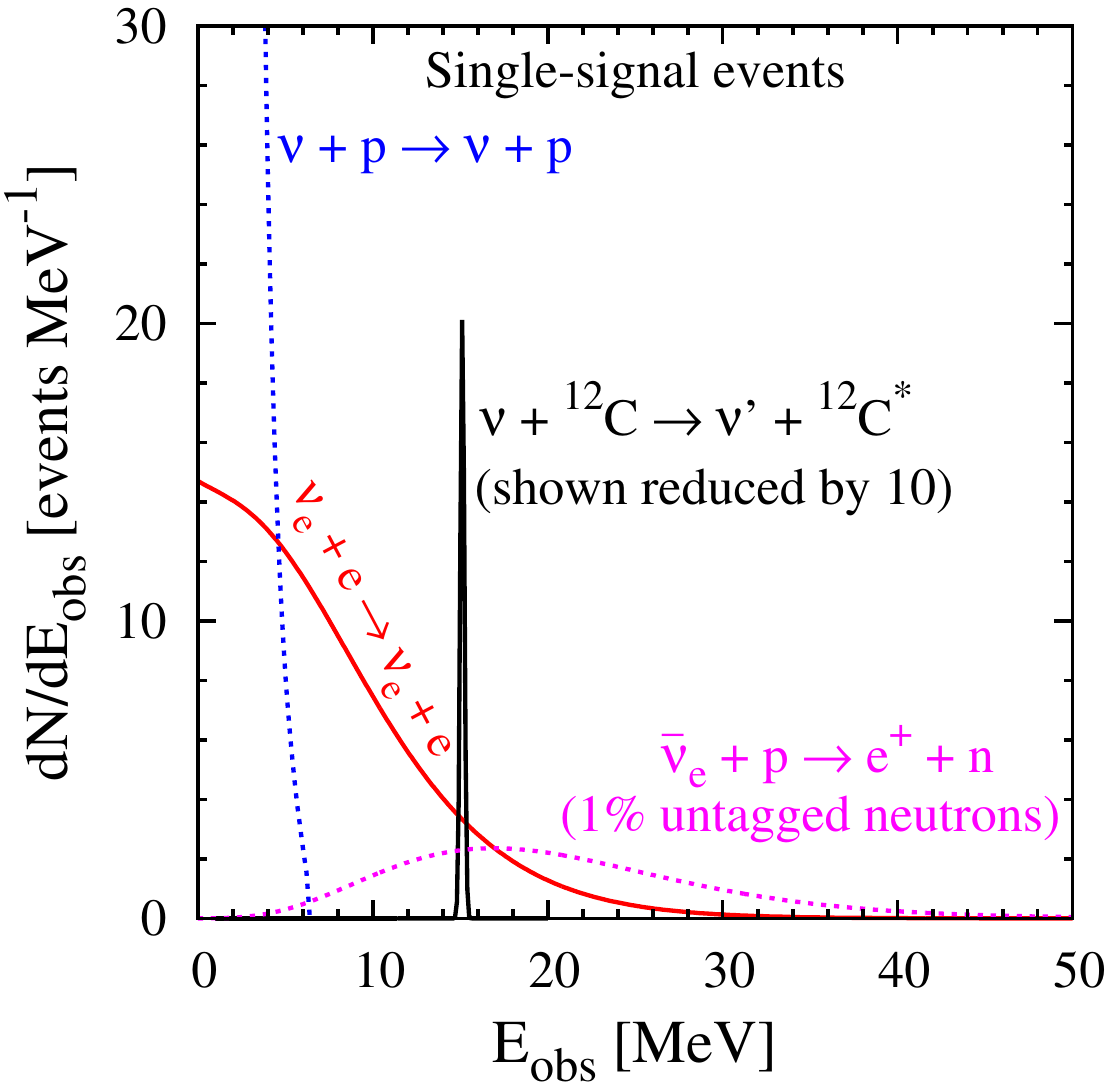}
\includegraphics[angle=0.0,width=0.48\textwidth]{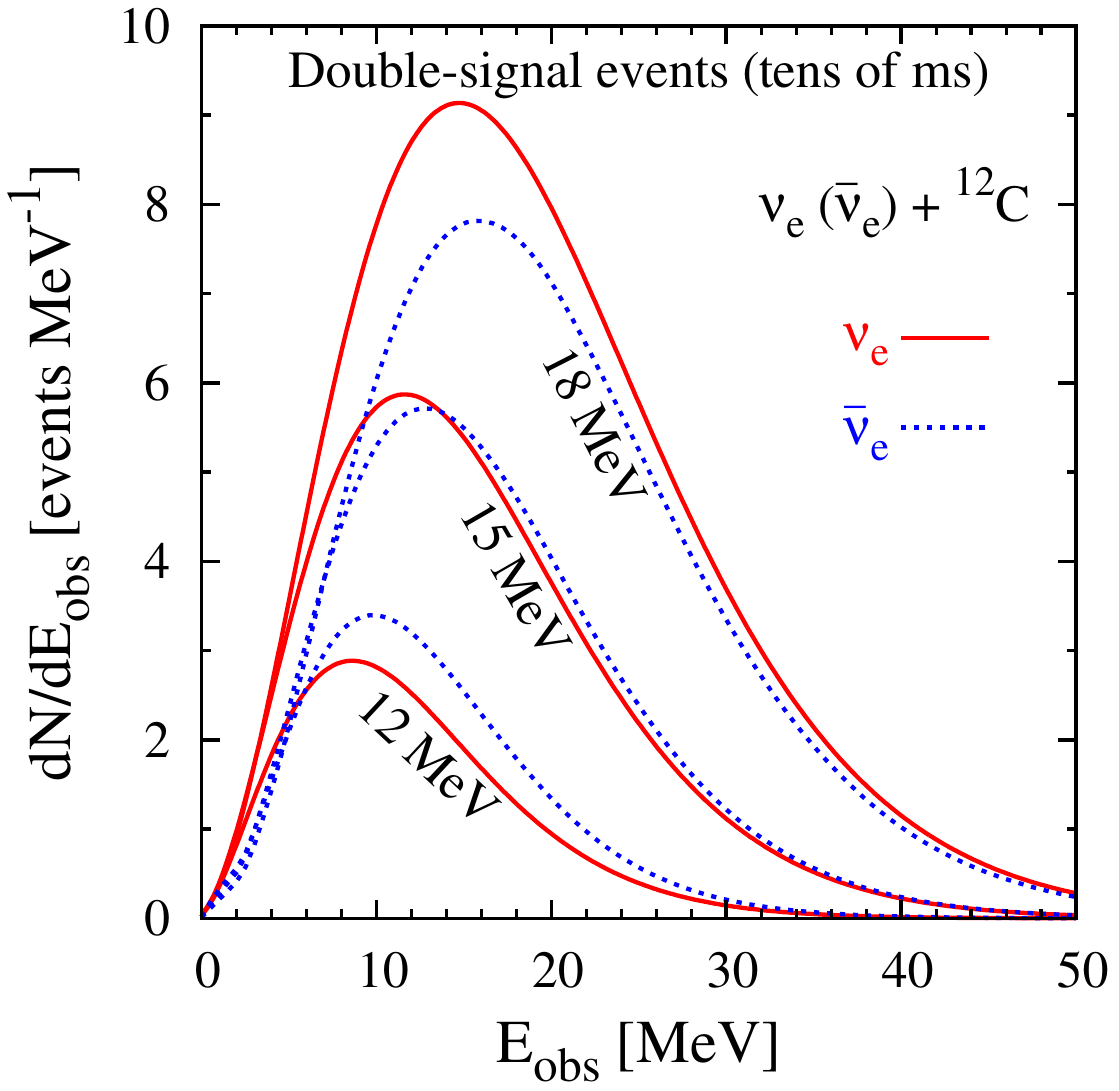}
\caption{Observed energy distribution in a 20 kton liquid scintillator detector for a Galactic supernova. In both panels, the total energy carried by each flavor of $\nu$ or $\bar{\nu}$ is 5 $\times$ 10$^{52}$ erg.  The ranges of the y-axes are different.  {\bf Left Panel:}  We assume $\langle E_{\nu_e} \rangle = 12$ MeV, $\langle E_{\bar{\nu}_e} \rangle = 15$ MeV and $\langle E_{\nu_x} \rangle = 18$ MeV.  We show the recoil spectrum of electrons in $\nu_e + e^-$ elastic scattering in solid red line.  The total recoil spectrum of the protons in $\nu + p$ elastic scattering is shown by the dotted blue line.  The recoil spectrum of the positron in the untagged inverse beta interaction is shown by the dotted magenta line, assuming 99\% capture efficiency of neutrons.  The monochromatic 15.11 MeV gamma-ray line due to $\nu$ + $^{12}$C neutral current interaction with $\langle E_{\nu} \rangle$ = 15.11 MeV is shown in solid black line, including the effects of energy resolution~\cite{Li:2014qca} and reduced in height by a factor of 10.  {\bf Right panel:}  We show the recoil spectrum of electrons and positrons from $\nu_e$ + $^{12}$C and $\bar{\nu}_e$ + $^{12}$C interactions.  For both, the spectra are shown for three values of $\langle E_{\nu} \rangle$: 12 MeV, 15 MeV and 18 MeV.  The electron and positron spectra are shown in solid red and dotted blue, respectively.}
\label{fig:neutrino interactions}
\vspace{0.1cm}
\end{figure*}

We assume that the fiducial volume of the detector for supernova neutrino detection is 20 kton.  The detector backgrounds are negligible above 0.2 MeV during a supernova burst.  We closely follow the parameters of the detector that are described in~\cite{Li:2014qca} throughout this work.

It will be important to distinguish between the final states from $\nu_e$ + $^{12}$C and $\bar{\nu}_e$ + $^{12}$C interactions.  Pulse shape distortion produced by positron can be used to distinguish between these interactions.  The resultant nuclei in these interactions have a half-life of 11 msec and 20 msec, and this time structure can help to distinguish between these interactions.  The slightly different end points of the positron and electron spectrum from $^{12}$N$_{\rm g.s.}$ and $^{12}$B$_{\rm g.s.}$ can also help.  As the $\bar{\nu}_e$ spectrum will be known to $\sim$ 2\% precision from the inverse beta interaction, it can be used to predict the $\bar{\nu}_e$ + $^{12}$C signal.

The left panel in Fig.~\ref{fig:neutrino interactions} shows the recoil spectra for neutrino electron scattering events for $\nu_e$.  The electron scattering events due to $\bar{\nu}_e$ and $\nu_x$ are further suppressed and not shown for clarity (see Ref.~\cite{Laha:2013hva}).  We also show the differential event rate due to $\nu$ + $p$ elastic scattering interactions, which is dominated by $\nu_x$.  We have taken the average energies of the various neutrino flavors as $\langle E_{\nu_e} \rangle$ = 12 MeV, $\langle E_{\bar{\nu}_e} \rangle$ = 15 MeV, and $\langle E_{\nu_x} \rangle$ = 18 MeV in this plot.  Other than neutrino-proton scattering, no other single-signal interactions can be reliably detected below 5 MeV.

In the right panel of Fig.~\ref{fig:neutrino interactions}, we plot the recoil spectra of the prompt charged lepton due to $\nu_e$/ $\bar{\nu}_e$+ $^{12}$C charged current interactions.  The resultant nuclei decay with known lifetimes.  We show the recoil spectra of the resultant electron (positron) from the $\nu_e$ ($\bar{\nu}_e$) + $^{12}$C interaction for three values of the $\nue$ ($\bar{\nu}_e$) average energy, $\langle E \rangle$ = 12 MeV, 15 MeV, and 18 MeV.  The strong dependence of the average energy on these interactions is clearly visible in this plot.

\section{Constraints on Supernova $\nu_e$ spectral properties}
\label{sec:SN nue detection_in_liquid_scintillator}

In this section, we discuss the constraints on $\nu_e$ spectral parameters that can be obtained in a liquid scintillator detector.  We will assume a range of average energies of the thermal $\nu_e$ spectrum to determine the impact of neutrino oscillations in determining the $\nu_e$ spectral parameters.  We will also show constraints for non-thermal $\nu_e$ spectrum.

\subsection{Calculated Detection Spectra}

Neutrino oscillations in a supernova occur via MSW mechanism and collective neutrino mixing.  There is uncertainty in the neutrino spectrum before oscillation.  Predicting the average energies of the neutrino spectrum after oscillation is difficult due to varying matter density inside the supernova.  To take into account these uncertainties, we consider two extreme cases of $\nu_e$ average energies at first.  Interpretation of our results on $\nue$ average energy and total energy in terms of the underlying neutrino parameters and matter potential is complicated and can be done with least uncertainty after the occurrence of a Galactic supernova.

Case (A): $\langle E_{\nu_e} \rangle \approx 12$ MeV and $\langle E_{\nu_x} \rangle \approx$ 15 -- 18 MeV.  In this case, there is no mixing between $\nu_e$ and $\nu_x$ and the hierarchy of average energies due to the late decoupling of $\nu_e$ is maintained.

Case (B): $\langle E_{\nu_e} \rangle \approx$ 15 -- 18 MeV, one flavor of $\nu_x$ has an average energy $\approx$ 12 MeV, and the other flavors of $\nu_x$ have an average energy $\approx$ 15 -- 18 MeV.  In this case, neutrino mixing has interchanged the average energy of $\nu_e$ and one of the $\nu_x$.  This case is easily distinguishable in a water Cherenkov detector with gadolinium due to the large number of $\nu_e \,+\, ^{16}$O scattering events~\cite{Laha:2013hva}.

Our initial aim is to constrain the $\nue$ spectral properties for case (A) and case (B).  Spectral properties of all other neutrino flavors can be constrained by other neutrino intereactions: inverse beta interactions will constrain the $\bar{\nu}_e$ spectral properties, and $\nu \,+\, p$ elastic scattering will constrain the $\nu_x$ spectral properties.

It is evident from Table~\ref{tab:yields} that the neutrino + $^{12}$C charged current interactions are very sensitive to the average energy of the incoming neutrino spectrum.  The strong energy dependence ensures that a strong constraint on the average energy of the neutrino spectrum will be deduced from this interaction.  When the average energy of the neutrino spectrum is low, the number of neutrino + $^{12}$C charged current interactions are low, and the constraint on the total energy carried by the neutrino flavor is weak.  For higher values of $\langle E_{\nu_e} \rangle$, the constraints on the total energy carried by $\nu_e$ are stronger due to larger number of events.

\begin{figure*}[t]
\centering
\includegraphics[angle=0.0,width=0.497\textwidth]{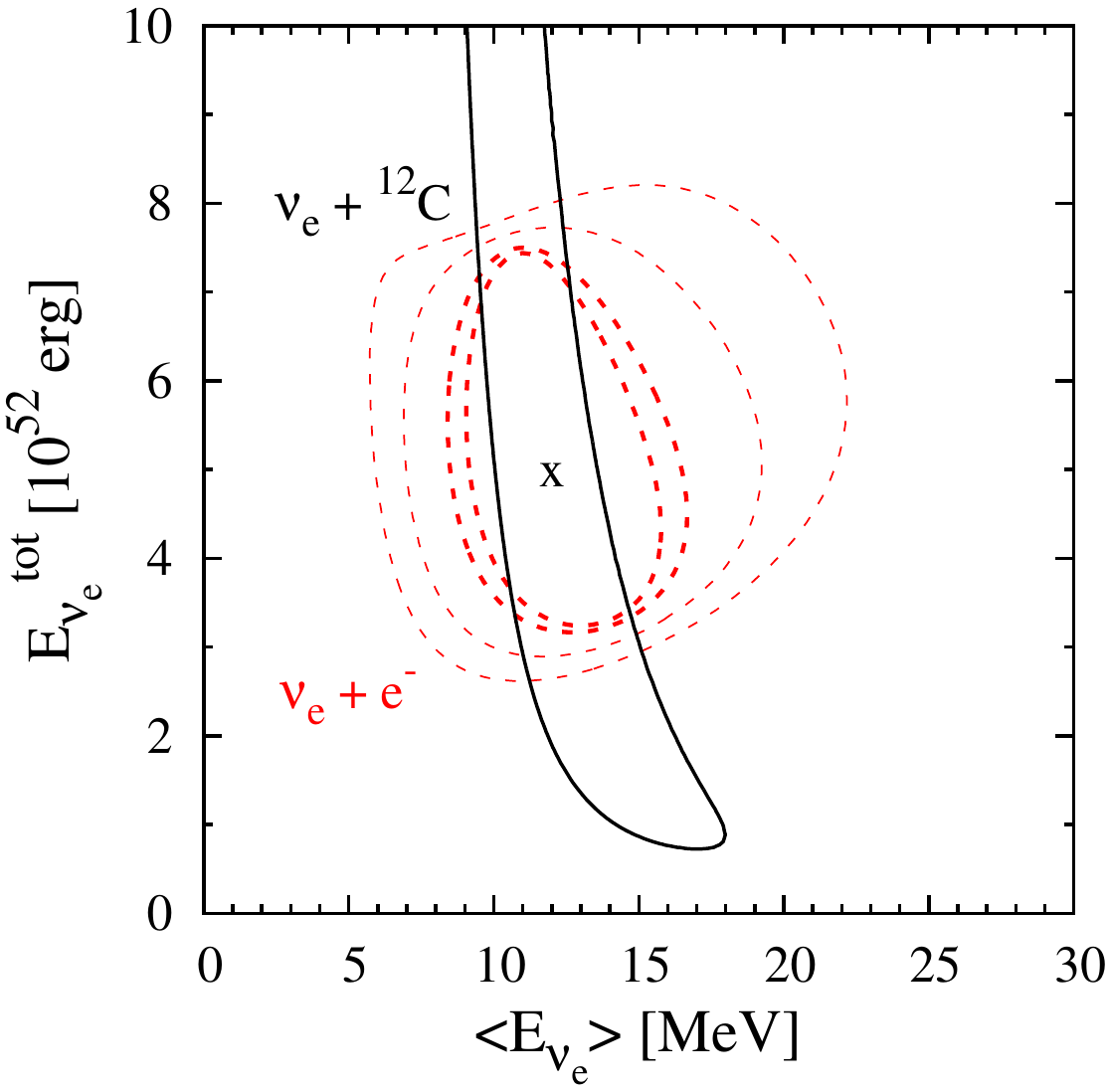}
\includegraphics[angle=0.0,width=0.497\textwidth]{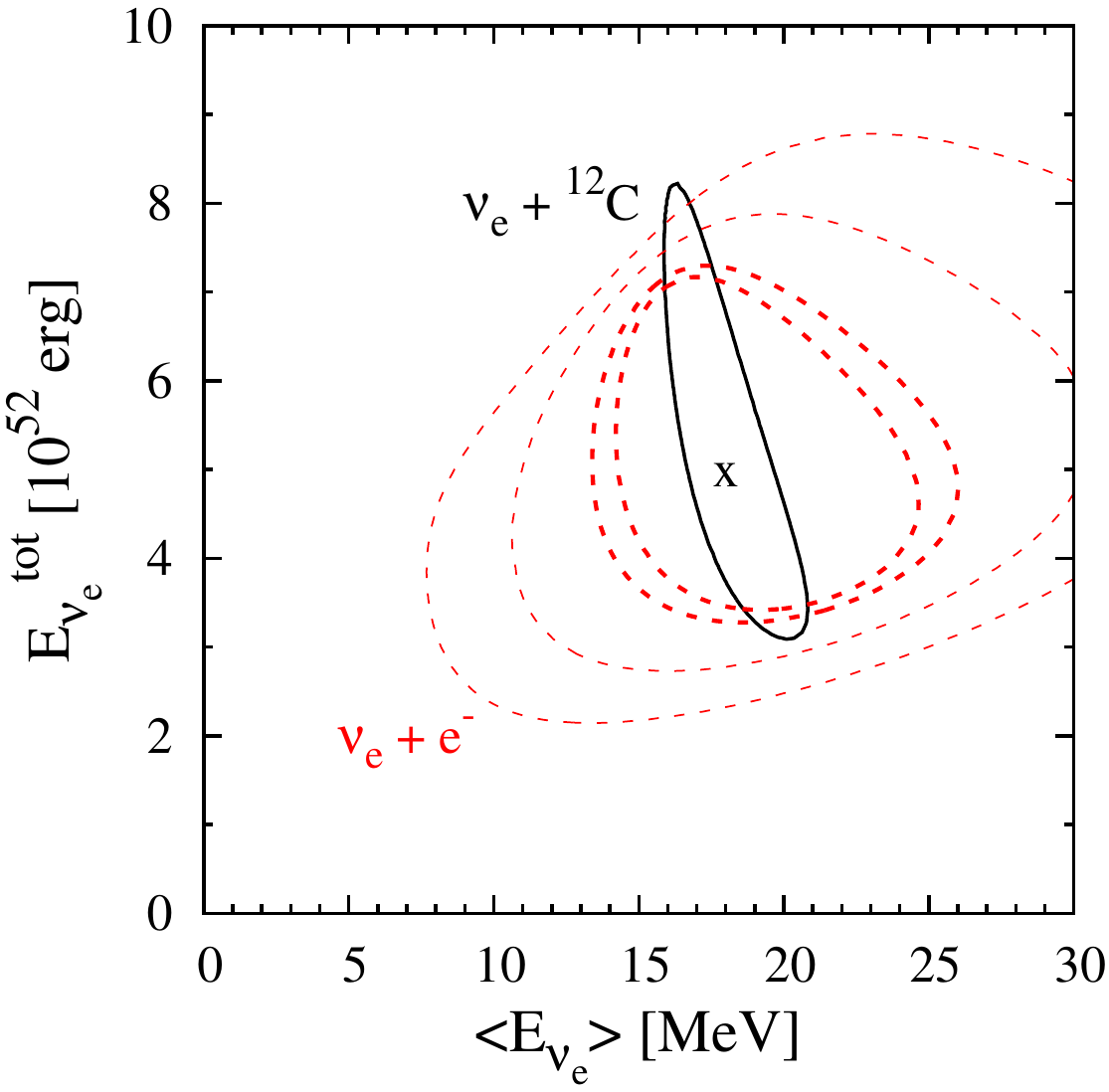}
\caption{Allowed parameter space (90\% C.L. $\Delta \chi ^2$ contours) for the $\nu_e$ spectrum parameters using the $\nu_e \,+\, e^-$ and $\nu_e \,+\, ^{12}$C channels.  We do not show the combined constraints which closely follow as would be expected visually.  The best fit point in both the panels are shown by x.  The total energy carried by all the neutrino flavors is 5$\times$ 10$^{52}$ erg.  In both cases we have closed contours on the neutrino spectral properties.  The four contours for the $\nu_e \,+\, e^-$ scattering events are for a neutron detection efficiency of 90\%, 95\%, 99\% and 100\% with decreasing distance from the point marked x respectively.  The capture efficiency of neutrons on proton will not have a large impact on the contour obtained from the $\nu_e \,+\, ^{12}$C interaction.    {\bf Left Panel:} $\langle E_{\nu_e} \rangle$ = 12 MeV and $\langle E_{\nu_x} \rangle$ = 18 MeV.  {\bf Right Panel:}  $\langle E_{\nu_e} \rangle$ = 18 MeV, i.e., one of the $\nu_x$ has oscillated into $\nu_e$.}
\label{fig:chisquared}
\end{figure*}

\begin{figure}[t]
\centering
\includegraphics[angle=0.0,width=0.48\textwidth]{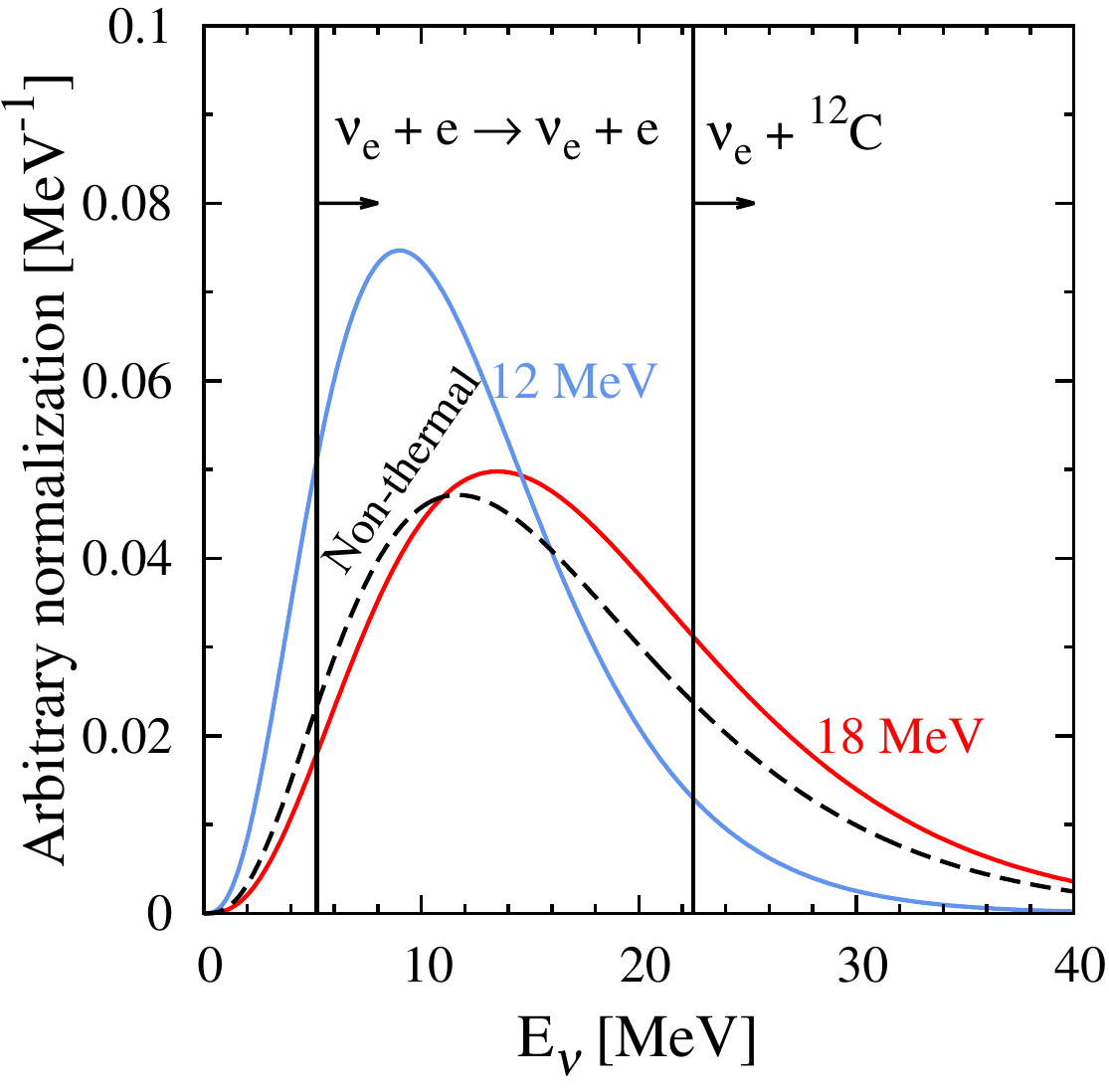}
\caption{The neutrino spectrum from a supernova and the part of the spectrum probed by $\nu_e \,+\, e^-$ elastic scattering and $\nu_e \,+\, ^{12}$C charged current interaction.  The neutrino spectrum (Eqn.\,\ref{eq:normalised spectra}) is shown for two different average energies of 12 MeV and 18 MeV in blue and red respectively.  The part of the neutrino spectrum probed by $\nu_e \,+\, e^-$ spectrum ($>$ 5.2 MeV) and by $\nu_e \,+\, ^{12}$C charged current interaction ($>$ 22.5 MeV) is shown by vertical lines.  We also show a non-thermal spectrum in dashed black which results from MSW mixing in the inverted hierarchy due to the two thermal spectra shown in the figure.}
\label{fig:spectrum probed}
\vspace{0.1cm}
\end{figure}

\begin{figure}[t]
\includegraphics[angle=0.0,width=0.47\textwidth]{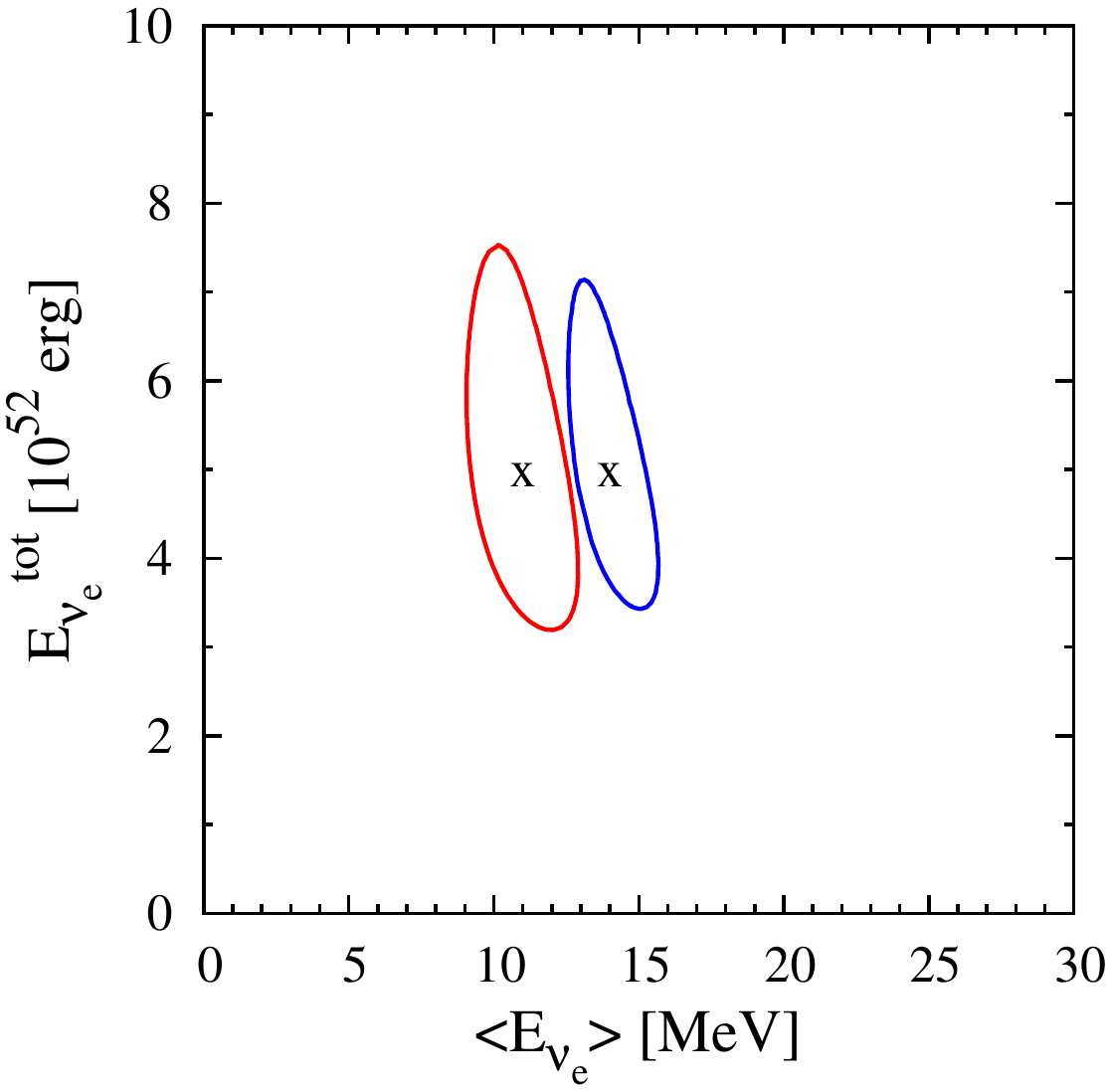}
\caption{Joint 90\% C.L. $\Delta \chi ^2$ contours for $\nu_e$ spectrum parameters determined from the $\nu_e \,+\, e^-$ and $\nu_e \,+\, ^{12}$C channels.  We distinguish two different cases of $\nu_e$ average energies here, 11 MeV and 14 MeV.  The total energy carried by the neutrino flavor is 5 $\times$ 10$^{52}$ erg.}
\label{fig:chisquaredjoint}
\end{figure}

The $\nu_e \,+\, ^{12}$C interactions can be detected with very high efficiency due to the double coincidence signal.  There is a very small probability that a fully tagged inverse beta signal will coincide with the double coincidence signal of the $\nu_e \,+\, ^{12}$C interaction.  The main reason for this is the factor $\sim$ 50 difference in the characteristic time of the two double coincidence signatures (207 $\mu$sec v.s. 11 msec).  The probability of a neutron capture on proton from inverse beta interaction to happen after 2 msec of the prompt positron signal is $\sim$ 6 $\times$ 10$^{-5}$.  

The double coincidence signal of $\nu_e \,+\, ^{12}$C can be confused with the double coincidence signal of $\bar{\nu}_e \,+\, ^{12}$C unless the positrons are tagged efficiently.  As mentioned earlier, positron detection via pulse shape distortion can be used to distinguish between these two interactions.  Since present technology can only detect those positrons which annihilate after formation of ortho-positronium, about 55\% of the $\bar{\nu}_e \,+\, ^{12}$C will pose a background to the signal from $\nu_e \,+\, ^{12}$C interaction.  The factor $\sim$ 2 difference in the decay times of the metastable nuclei can be further used to distinguish between these interactions.  For conservativeness, we have assumed that all the $\bar{\nu}_e \,+\, ^{12}$C interactions will be a background to the $\nu_e \,+\, ^{12}$C signal.  Detection of $\sim$ 45\% will help in a slight improvement of the constraints on total energy, whereas the improvement on average energy will be negligible.  A future perfect discrimination of positrons from electrons can only improve this constraint at most by a factor of $\sim$ $\sqrt{2}$.

\begin{figure}[t]
\includegraphics[angle=0.0,width=0.47\textwidth]{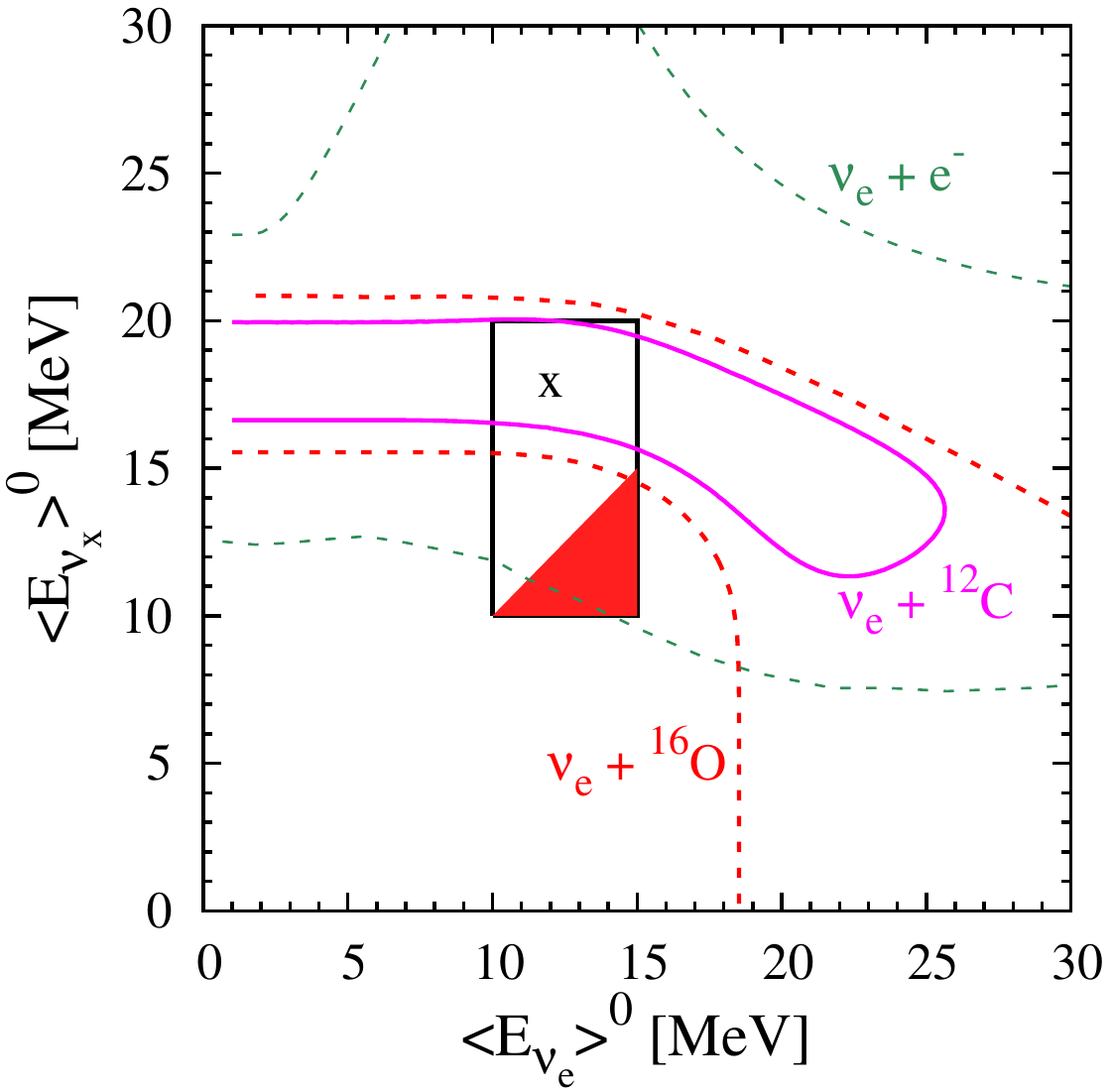}
\caption{90\% C.L. $\Delta \chi ^2$ contours for a non-thermal spectrum of $\nu_e$ in the plane of initial $\nu_e$ and $\nu_x$ average energies.  We take the example of MSW mixing in inverted hierarchy --- the incoming $\nu_e$ spectrum is given by an appropriate mixture of $\nu_e$ and $\nu_x$ spectrum (see black dashed curve in Fig.\,\ref{fig:spectrum probed}).  We fix the total energy carried by the $\nu_e$ and $\nu_x$ flavor as 5 $\times$ 10$^{52}$ erg.  The best fit point for the initial average energies for $\nu_e$ and $\nu_x$ is given by x.  The contour obtained from the $\nu_e \,+\, ^{16}$O and $\nu_e \,+\, ^{12}$C interaction in gadolinium loaded Super-Kamiokande and future large liquid scintillator detector is given by the red dashed line and magenta line respectively.  The constraint obtained from $\nu_e$ + $e^-$ scattering in a gadolinium loaded Super-Kamiokande detector is shown in dashed green.  We have assumed a neutron capture efficiency of 90\% in gadolinium loaded Super-Kamiokande to obtain this result.  The rectangle shows the theoretically expected region of $\langle E_{\nu_e} \rangle ^0$ and $\langle E_{\nu_x} \rangle ^0$.  The lower part of the rectangle shaded in red is excluded since $\langle E_{\nu_e} \rangle ^0 > \langle E_{\nu_x} \rangle ^0$ in that region, which is contrary to theoretical expectations.  }
\label{fig:MSWIH}
\end{figure}

\subsection{Neutrino Spectral Parameter fit}

We have performed the usual $\chi^2$ analysis using a Poissonian likelihood function.~\cite{Agarwalla:2006vf}.  We take the systematic uncertainty on background and signal as 5\% and 10\% respectively.  The short duration of a supernova burst means that the background is much less uncertain.  The systematic uncertainty on the background and signal can only be properly quantified by the experimental collaboration.  Due to $\sim$ 10\% uncertainties throughout the work, we use these values as indicative.  We only fit for the $\nu_e$ spectral parameters as the spectral parameters of other flavors will be known from measurements in other detection channels.

We calculate the $\Delta \chi^2$ of the various best fit values.  We use $\Delta \chi^2$ = 4.6 for two degrees of freedom to get the 90$\%$ C.L. regions.  We have made various assumptions of the order of 10\% throughout this work, so the contours should be understood with about 10\% uncertainty.

Since the capture efficiency of neutrons on protons are unknown, we will show the contours for four different values of this efficiency: 90\%, 95\%, 99\% and 100\%.  The contours obtained from $\nu_e \,+\, e^-$ scattering depend quite strongly on this capture efficiency.  This is obvious as in the absence of neutron tagging, the prompt positrons from inverse beta interactions form a background to the search of recoil electrons from $\nu_e \,+\, e^-$ scattering.  

The detection efficiency of neutrons have a negligible effect on the contours obtained from $\nu_e \,+\, ^{12}$C scattering.  The double coincidence signal of $\nu_e \,+\, ^{12}$C interaction has a very different characteristic time compared to the inverse beta interaction.  The inverse beta interactions with untagged neutrons are single-signal events and hence will not be confused with the double-signal event characteristic of $\nu_e \,+\, ^{12}$C.

We show the likely constraints for case (A) in the left column of Fig.~\ref{fig:chisquared}.  The best fit parameters in this case are $\langle E_{\nu_e} \rangle$ = 12 MeV and $E_{\nu_e} ^{\rm tot}$ = 5 $\times$ 10$^{52}$ erg and is shown by x.  We show the contours for four different values of capture efficiency of neutrons : 90\%, 95\%, 99\% and 100\%.  For this low value of $\langle E_{\nu_e} \rangle$, the constraint on the total energy from the $\nu_e \,+\, ^{12}$C charged current interaction is very weak.  Due to the absence of directionality, the constraint from the $\nu_e \,+\, e^-$ elastic scattering is weaker than the constraint obtained from a water Cherenkov detector.  The joint contour of the two detection channel is similar to that one can obtain by eye. The total energy carried by $\nu_e$ can determined to $\sim$ 40\% precision and the average energy of $\nu_e$ can be determined to $\sim$ 25\% precision in this case.  

When there is oscillation between $\nu_e$ and one of the flavors of $\nu_x$, the constraints from $\nu_e \,+\, ^{12}$C charged current interactions become strong.  These are shown in the right panel of Fig.~\ref{fig:chisquared}.  The best fit parameters shown by x are $\langle E_{\nu_e} \rangle$ = 18 MeV and $E_{\nu_e} ^{\rm tot}$ = 5 $\times$ 10$^{52}$ erg.  The increase in the average energy hardly increases the number of events from $\nu_e \,+\, e^-$ elastic scattering and hence the constraints on average energies from this elastic scattering is extremely weak.  The large number of $\nue \,+\, ^{12}$C charged current interactions ensure that the constraints obtained from the double coincidence signal is extremely strong.  In this case, the total energy carried by $\nue$ is determined to $\sim$ 40\% precision and the average energy is determined to $\sim$ 10\% precision.

As is easily seen from the figure, the sharp energy dependence of the $\nu_e \,+\, ^{12}$C interaction will help separate the two cases when the average neutrino energies are far apart.  This is an important result and it will help us understand supernova much better.  Both the $\nu_e \,+\, e^-$ scattering and the $\nu_e \,+\, ^{12}$C scattering give weak constraints on the total energy carried by the $\nu_e$ flavor.

For the joint contours, the stronger constraint on the $\langle E_{\nu_e} \rangle$ is always obtained from the $\nu_e \,+\, ^{12}$C interaction.  The stronger constraint on the $E^{\rm tot}_{\nu_e}$ is generally obtained from the $\nu_e \,+\, e^-$ interaction, unless the $\langle E_{\nu_e} \rangle$ is high such that there are a large number of $\nu_e \,+\, ^{12}$C interactions.  This complementary information carried by both the interactions should be fully utilized to obtain the best information from these two important channels for $\nu_e$ detection in liquid scintillator detector.

The part of the neutrino spectrum probed by $\nu_e \,+\, e^-$ elastic scattering and $\nu_e \,+\, ^{12}$C charged current interaction is quite different and is shown in Fig.\,\ref{fig:spectrum probed}.  The detection threshold is taken to be 5 MeV, and hence the part of the neutrino spectrum probed by $\nu_e \,+\, e^-$ and $\nu_e \,+\, ^{12}$C charged current interaction is $\gtrsim$ 5.2 MeV and $\gtrsim$ 22.5 MeV respectively.  The $\nu_e \,+\, ^{12}$C charged current interaction is only sensitive to higher neutrino energies, and hence the strong dependence on the neutrino average energy.  A slight change in the neutrino average energy will change the tail of the spectrum significantly and this explains the strong constraint on $\langle E_{\nu_e} \rangle$ obtained from the $\nu_e \,+\, ^{12}$C interaction.
 
We now discuss cases in which the average energy of the incoming $\nu_e$ spectrum have less hierarchy.  Modern computer simulations of supernova explosion typically show a much closer range of the average energy of neutrinos of different flavors~\cite{Mueller:2012ak}, $\langle E_{\nu_e} \rangle \approx$ 11 MeV, and $\langle E_{\bar{\nu}_e} \rangle \approx \langle E_{\nu_x} \rangle \approx$ 14 MeV.  In our companion paper~\cite{Laha:2013hva}, we had taken these values to be 11 MeV and 15 MeV.  The potential better discrimination possible in a large liquid scintillator detector motivates us to consider this slightly more challenging situation.  Due to the proximity of the average energies, it is much more difficult to determine whether $\nu_e \leftrightarrow \nu_x$ oscillations have taken place.  

We show the 90\% joint contours (using both the neutrino electron elastic scattering and neutrino carbon charged current interaction) for this situation in Fig.~\ref{fig:chisquaredjoint}.  The efficiency of neutron detection is taken to be 100\% in this plot.  The total energy carried by each neutrino flavor is taken to be 5$\times$ 10$^{52}$ erg.  The strong energy dependence of the neutrino carbon charged current interactions help us separate the two cases of mildly differing average energies.  This will be an important physics motivation for large liquid scintillator detectors.  The height of both these contours depend on the constraint from $\nu_e$ + e$^-$ scattering and the $\nu_e$ + $^{12}$C charged current scattering controls the width of these contours.  Having a less than perfect capture efficiency of neutrons will degrade this constraint on the total energy but will not affect the separation in the average energies.

\subsection{Non-thermal $\nu_e$ spectrum}

As a final illustration of the discriminating power of large liquid scintillator detectors, we try to reconstruct the spectral parameters for a non-thermal spectrum of incoming $\nu_e$.  We only consider MSW oscillations for this example~\cite{Dighe:1999bi,Dighe:2007ks}.  Collective oscillations will generally complicate the situation further and a dedicated study is required for that purpose~\cite{Choubey:2010up,Dasgupta:2010gr}.  The normal hierarchy scenario in MSW mixing is similar to the ones we considered earlier in this work: the final $\nu_e$ spectrum emitted from the supernova is the initial $\nu_x$ spectrum.

In the inverted hierarchy for MSW mixing, the final $\nu_e$ spectrum is a mixture of the initial $\nu_e$ spectrum (mixing probability = sin$^2$ $\theta_{\odot}$) and the initial $\nu_x$ spectrum (mixing probability = 1- sin$^2$ $\theta_{\odot}$).  If the original $\nu_e$ and $\nu_x$ average energy is 12 MeV and 18 MeV respectively, then the final spectrum after MSW mixing is shown by the black dashed curve in Fig.\,\ref{fig:spectrum probed}.

The total spectrum is then determined by 4 parameters:  the total and average energy carried by the $\nu_e$ and $\nu_x$ flavor.  We can relate the total energies carried by $\nu_e$ and $\nu_x$ flavor since we will know the total binding energy of the supernova and the total energy carried by $\bar{\nu}_e$ (from inverse beta interactions).  This reduces the number of free parameters to three.

Due to the non-thermal spectrum and the three free parameters, a complete scan of the parameter space is complicated and require a dedicated study.  We will show the allowed contours for the initial average energies carried by the $\nu_e$ ($\langle E_{\nu_e}\rangle ^0$) and $\nu_x$ ($\langle E_{\nu_x}\rangle ^0$) flavor assuming that the total energy carried by the neutrino flavors is 5 $\times$ 10$^{52}$ erg.  For a full scan, we have to vary the value of the total energy carried by the neutrino flavors.  We fix it to a fiducial value here for a simple pedagogical example.

Since the spectrum has a non-linear dependence on the average energies, we only show the constraints on $\langle E_{\nu_e}\rangle ^0$ and $\langle E_{\nu_x}\rangle ^0$ in Fig.\,\ref{fig:MSWIH}.  The best fit point, shown by x, is given by $\langle E_{\nu_e}\rangle ^0$ = 12 MeV and $\langle E_{\nu_x}\rangle ^0$ = 18 MeV.  The rectangle encloses the range 10 MeV $\leq$ $\langle E_{\nu_e}\rangle ^0$ $\leq$ 15 MeV, and 10 MeV $\leq$ $\langle E_{\nu_x}\rangle ^0$ $\leq$ 20 MeV, which is approximately the theoretically favored region according to supernova simulations.  The lower part of the rectangle shaded in red is disfavored as we expect  $\langle E_{\nu_x}\rangle ^0$ $>$ $\langle E_{\nu_e}\rangle ^0$.

We show the constraint from both $\nu_e \,+\, ^{12}$C and $\nu_e \,+\, ^{16}$O interaction in a near-future large liquid scintillator detector and gadolinium loaded Super-Kamiokande respectively.  We also show the constraint from $\nu_e$ + e$^-$ interactions in gadolinium loaded Super-Kamiokande detector.  We assume that the neutron capture efficiency on gadolinium in Super-Kamiokande and on protons (and carbon) in liquid scintillator detector is 90\% and 100\% respectively.  As is evident from the black dashed curve in Fig.\,\ref{fig:spectrum probed}, the non-thermal spectrum at high energies is mostly due to the flavor with higher average energy.  This explains the nearly horizontal behavior of the contours for the neutrino nucleus interactions.  The constraint on the flavor with higher average energy is around 10\% and which is approximately equal to the constraint that one can achieve from $\nu$ + $p$ interaction as well.  The $\nu_e$ + e$^-$ elastic scattering is nearly independent of the average energy and hence the constraint from that interaction is very broad and uninformative.

The constraint obtained from the $\nu_e \,+\, ^{16}$O interaction in gadolinium loaded Super-Kamiokande is slightly weaker due to the larger number of background events.  In this plot, a contour with a smaller area denotes better discriminating power.  The constrained area is approximately 46\% of the theoretically favored region.  Lowering the threshold of the neutrino nucleus interaction will further improve these constraints.  Our results should encourage the experimentalists to optimise their cuts to detect $\nu_e$ + $^{12}$C interaction in a very efficient manner, since this interaction will give us the strongest constraint on the average energies for a non-thermal spectrum.
 
We can compare the results obtained in the supernova $\nu_e$ detection in Super-Kamiokande loaded with gadolinium~\cite{Laha:2013hva} with the results obtained in this work.  It is important to detect supernova neutrinos in many different detectors as earth matter effects can be understood if the detectors are far apart.  Detection of neutrinos in two different detection methods will also improve the reliability of the detection.  The presence of directionality in water Cherenkov detectors help in finding the direction of the Galactic supernova; this is not possible in a liquid scintillator detector.  The charged current neutrino interaction on $^{16}$O is more uncertain than the corresponding interaction on $^{12}$C as the latter has been measured.  One advantage for water Cherenkov detectors is that they can be designed to be a factor of $\sim$20 larger than Super-Kamiokande with the current technology and there is active research in that direction~\cite{Kearns:2013lea}.  Having such a large water Cherenkov detector loaded with gadolinium will improve the uncertainties of the $\nu_e$ spectral parameters by a factor of $\sim$ 5~\cite{Laha:2013hva}.  Currently there are no plans to build a liquid scintillator detector of the size of Hyper-Kamiokande.  Detection of supernova $\nu_e$ in these two different detectors will involve different systematics and this will help in improving our knowledge of supernova physics.

\section{Conclusions}
\label{sec:conclusion_in_liquid_scintillator}

Detecting supernova neutrinos is important for astrophysics and neutrino physics~\cite{Janka:1995bx, Langanke:2002ab, Mezzacappa:2005ju, Burrows:2006ci, Janka:2006fh, Woosley:2006ie, Burrows:2012ew, Janka:2012wk,Ott:2008wt,Kotake:2011yv,Ando:2012hna,Yuksel:2012zy,Woosley:2006fn,Woosley:1989bd,Thielemann:2001rn,Heger:2003mm,Woosley:2007as,Kistler:2012as}.  In spite of detecting numerous supernovae via electromagnetic signal, we still do not completely understand a supernova.  One of the main reasons behind this is that the total energy budget of the supernova is completely dominated by neutrinos.  The only way to completely understand a supernova is to detect all flavors of neutrinos emitted by it.  No one type of detection technique is completely efficient in detecting all the supernova neutrino flavors and hence it is essential to look into various different detector types to improve our knowledge of supernova neutrinos.   It is important to detect supernova neutrinos in different detectors to increase detectability of the signal as other detectors might not be working during that short period of time.

Detection of supernova $\nu_e$ in a gadolinium loaded water Cherenkov detector was investigated in~\cite{Laha:2013hva}.  In this companion work, we determine the feasibility of detecting supernova $\nu_e$ in near-future large liquid scintillator detector.  The detection technique in these two different detectors are different and we gain more knowledge in detecting supernova $\nu_e$ in both these different types of detectors.  The main interactions of $\nu_e$ in a liquid scintillator detector are its elastic scattering with electrons and charged current interaction with $^{12}$C nuclei.  The former interaction is almost independent of the average energy of the $\nu_e$ flavor, but the later interaction will give us a precise measure of the average energy of the $\nu_e$.

As in any supernova neutrino detection technique, the largest background for the $\nu_e$ detection is the inverse beta interaction caused by the supernova $\bar{\nu}_e$.  Neutron capture on proton and carbon will help reduce this enormous background.  The charged current interaction of $\nu_e$ on $^{12}$C nuclei can be distinguished by the double coincidence signal of this interaction.  There is a possibility that this interaction might be confused with charged current interaction of $\bar{\nu}_e$ with $^{12}$C; identification of positron via pulse shape distortion and the separation of the events due to the different decay times of the metastable nuclei will help in their distinction.  It should also be possible to statistically subtract the $\bar{\nu}_e$ charged current interaction as inverse beta interactions will give us a very precise measure of the $\bar{\nu}_e$ spectrum.  The neutral current interaction of neutrinos on $^{12}$C will also help in deterring the total binding energy of the supernova independent of oscillation physics.

The main results of this work is shown in Figs.\,\ref{fig:chisquared}, \ref{fig:chisquaredjoint} and \ref{fig:MSWIH}.  From Fig.\,\ref{fig:chisquared}, we find that the total energy and the average energy carried by the $\nu_e$ flavor will be known to $\lesssim$ 25\% and 40\% precision respectively.  More importantly, 20 kton liquid scintillator detectors can distinguish between whether the average energy of $\nu_e$ is 11 MeV or 14 MeV (Fig.\,\ref{fig:chisquaredjoint}).  This ability to distinguish between different $\nu_e$ average energies make liquid scintillator detector an important tool in understanding supernova.  The interesting constraint on the initial average energies of the $\nu_e$ and $\nu_x$ flavors for a non-thermal spectrum of $\nu_e$ due to MSW mixing in inverted hierarchy is shown in Fig.\,\ref{fig:MSWIH}.

All the constraints presented in this work can potentially be improved by a factor of $\sim$ $\sqrt{2}$ if data from two independent and similar sized detectors like JUNO and RENO-50 are combined.  If a larger detector like LENA detects supernova neutrinos, then LENA by itself should improve the presented constraints by a factor of $\sim$ 1.6.  Supernova $\nu_e$ can be detected with high precision in large liquid Argon detectors which are not yet built~\cite{GilBotella:2003sz,GilBotella:2004bv}. 

It is interesting to note that a liquid scintillator detector can detect all the different flavors of supernova neutrinos: $\bar{\nu}_e$ through inverse beta interaction, $\nu_x$ through the elastic scattering with protons and $\nu_e$ through its elastic interaction with electrons and charged current interaction with $^{12}$C.  Due to the serious interest in large liquid scintillator detectors~\cite{Wurm:2011zn,Li:2014qca,RENO-50:2013,Mollenberg:2014pwa,Mollenberg:2014mfa}, it is important for people to look into the details of supernova neutrino detection in these detectors.  We hope that our work will encourage the experimentalists to optimize their cuts to detect the neutrino signal from a Galactic supernova.


\section*{Acknowledgments} 
We thank Basudeb Dasgupta, Shunsaku Horiuchi, Patrick Huber, Leonidas Kalousis, Mattew Kistler, Jonathan Link, Shirley Li, Camillo Mariani and Yifang Wang for discussions.   RL was supported in the initial part of the work by NSF Grant PHY-1101216 awarded to JFB.  JFB was supported by NSF Grant PHY-1101216 and PHY-1404311.  S.K.A. acknowledges the support from DST/INSPIRE Research Grant [IFA-PH-12], Department of Science and Technology, India.

\bibliographystyle{kp}
\interlinepenalty=10000
\tolerance=100
\bibliography{Bibliography/references}

\end{document}